\documentclass[conference]{IEEEtran}
\IEEEoverridecommandlockouts
\usepackage{cite}
\usepackage{amsmath,amssymb,amsfonts}
\usepackage{algorithmic}
\usepackage{graphicx}
\usepackage{textcomp}
\usepackage{xcolor}
\usepackage{bm}
\usepackage{url}
\usepackage{booktabs, multicol, multirow}
\usepackage{subfigure}
\usepackage[T1]{fontenc}
\usepackage{aecompl}
\usepackage{flushend}
\def\BibTeX{{\rm B\kern-.05em{\sc i\kern-.025em b}\kern-.08em
    T\kern-.1667em\lower.7ex\hbox{E}\kern-.125emX}}

\begin{document}

\title{Detecting Code Clones with Graph Neural Network and Flow-Augmented Abstract Syntax Tree}

\author{
    \IEEEauthorblockN{Wenhan Wang\IEEEauthorrefmark{1}\IEEEauthorrefmark{2}, Ge Li\IEEEauthorrefmark{1}\IEEEauthorrefmark{2}\IEEEauthorrefmark{4}, Bo Ma\IEEEauthorrefmark{1}\IEEEauthorrefmark{2}, Xin Xia\IEEEauthorrefmark{3}, Zhi Jin\IEEEauthorrefmark{1}\IEEEauthorrefmark{2}\IEEEauthorrefmark{4}}
    \IEEEauthorblockA{\IEEEauthorrefmark{1}Key laboratory of High Confidence Software Technologies (Peking University), Ministry of Education
    }
    \IEEEauthorblockA{\IEEEauthorrefmark{2} Institute of Software, EECS, Peking University, Beijing, China\\\{wwhjacob, lige, 1700012844, zhijin\}@pku.edu.cn}
    \IEEEauthorblockA{\IEEEauthorrefmark{3} Faculty of Information Technology, Monash University, Melbourne, Australia
    \\xin.xia@monash.edu}

}

\maketitle
\renewcommand{\thefootnote}{}
\footnotetext[1]{\IEEEauthorrefmark{4} Corresponding Authors}
\begin{abstract}
Code clones are semantically similar code fragments pairs that are syntactically similar or different. Detection of code clones can help to reduce the cost of software maintenance and prevent bugs. Numerous approaches of detecting code clones have been proposed previously, but most of them focus on detecting syntactic clones and do not work well on semantic clones with different syntactic features. To detect semantic clones, researchers have tried to adopt deep learning for code clone detection to automatically learn latent semantic features from data. Especially, to leverage grammar information, several approaches used abstract syntax trees (AST) as input and achieved significant progress on code clone benchmarks in various programming languages. However, these AST-based approaches still can not fully leverage the structural information of code fragments, especially semantic information such as control flow and data flow. To leverage control and data flow information, in this paper, we build a graph representation of programs called flow-augmented abstract syntax tree (FA-AST). We construct FA-AST by augmenting original ASTs with explicit control and data flow edges. Then we apply two different types of graph neural networks (GNN) on FA-AST to measure the similarity of code pairs. As far as we have concerned, we are the first to apply graph neural networks on the domain of code clone detection. 

We apply our FA-AST and graph neural networks on two Java datasets: Google Code Jam and BigCloneBench. Our approach outperforms the state-of-the-art approaches on both Google Code Jam and BigCloneBench tasks.
\end{abstract}


\section{Introduction}
Code clone detection aims to measure the similarity between two code snippets. Commonly, there are two kinds of similarities within code clones: syntactic similarity and semantic similarity. Syntactic similarity are often introduced when programmers copy a code fragment and then pasting to another location, while semantic similarity occurs when developers try to implement a certain functionality which is identical or similar to an existing code fragment.

To better study the effectiveness of clone detectors on different types of code similarities, researchers started to systematically categorize code clones into multiple classes. One common taxonomy proposed by \cite{roy2007survey} group code clones into four types. The first three types of clones can be concluded as syntactic similarities, while type-4 clone can be seen as semantic similarity. As type-4 clones include clones that are highly
dissimilar syntactically, it is the hardest clone type to detect for most clone detection approaches. Code syntactical similarity has already been well-studied, while in recent years researchers have started to focus on detecting code semantic similarity. Along with the advances of deep neural networks, several deep learning-based approaches tried to capture the semantic similarities through learning from data. Most of these approaches include two steps: use neural networks to calculate a vector representation for each code fragment, then calculate the similarity between two code vector representations to detect clones. To leverage the explicit structural information in programs, these approaches often use abstract syntax tree (AST) as the input of their models \cite{white2016deep,wei2017supervised,zhang2019novel}. A typical example of these approaches is CDLH \cite{wei2017supervised}, which encode code fragments by directly applying Tree-LSTM \cite{tai2015improved} on binarized ASTs. Although AST can reflect the rich structural information for program syntax, it does not contain some semantic information such as control flow and data flow. 

To exploit explicit control flow information, some researchers use control flow graphs (CFG) to detect code clones. For example, DeepSim \cite{zhao2018deepsim} extracts semantic features from CFGs to build semantic matrices for clone detection. But CFG still lacks data flow information. Furthermore, most CFGs only contain control flows between code blocks and exclude the low-level syntactic structure within code blocks. Another drawback of CFGs is that in some programming languages, CFGs are much harder to obtain than ASTs.

In this paper, we aim to build a graph representation form of programs which can reflect both syntactical information and semantical information from ASTs. In order to detect code clones with the graphs we have built, we propose a new approach that uses graph neural networks (GNN) to detect code clones. Our approach mainly includes three steps: First, create graph representation for programs. Second, calculate vector representations for code fragments using graph neural networks. Third, measure code similarity by measuring the similarity of code vector representations. To fully leverage the control flow and data flow information in programs, we construct an AST-based graph representation of programs which we call it flow-augmented AST (FA-AST). FA-AST is constructed by adding various types of edges representing different types of control and data flow to ASTs. After we build the FA-AST for code fragments, we apply two different GNN models, gated graph neural network (GGNN) \cite{li2015gated} and graph matching network (GMN) \cite{li2019graph} on FA-ASTs to learn the feature vectors for code fragments. The first model separately computes vector representations for different code fragments, while the latter one jointly computes vector representations for a code pair. After we get the vector representations for a code pair, by measuring the similarity between them, we can determine whether these two code fragments belong to a clone.

In this paper, we build FA-AST for Java programs and evaluate FA-AST and graph neural networks on two code clone datasets: Google Code Jam dataset collected by \cite{zhao2018deepsim} and the widely used clone detection benchmark BigCloneBench \cite{svajlenko2014towards}. The results show that our approach outperforms most existing clone detection approaches, especially several AST-based deep learning approaches including RtvNN \cite{white2016deep}, CDLH \cite{wei2017supervised} and ASTNN \cite{zhang2019novel}.

The main contributions of this paper are as follows:

1) To the best of our knowledge, we are the first to apply graph neural networks on code clone detection. We adopt two different types of graph neural networks and analyze the difference between their performances.

2) We design a novel graph representation form FA-AST for Java programs that leverage both control and data flow of programs. Our graph representation is purely AST-based and can easily be extended to other programming languages.

3) We evaluate our approach on two datasets: Google Code Jam and BigCloneBench. Our approach performs comparable to state-of-the-art approaches on BigCloneBench and outperforms state-of-the-art approaches on Google Code Jam.

The remainder of this paper is structured as follows: Section II introduces the background knowledge. Section III defines the problem we aim to solve. We present the details of our approach in Section IV. We evaluate our approach and analyze its performance in Section V. In Section VI we discuss some findings in our experiments and some possible future improvements to our work. Section VII lists the related works. Finally, in Section VIII, we make a conclusion about our work.
\section{Background}
In this section we will introduce the background knowledge of code clone detection and graph neural networks (GNNs).
\subsection{Code Clone Detection}
According to \cite{roy2007survey}, code clones can be categorized into the following four types:

Type-1 (T1): Syntactically identical code fragments, except for differences in white space and comments.

Type-2 (T2): In addition to Type-1 clone differences, syntactically identical code fragments, except for differences in identifier names and literal values.

Type-3 (T3): In addition to Type-1 and Type-2 clone differences, syntactically similar code fragments that differ at the statement level. These fragments can have statements added, modified, and/or removed with respect to each other.

Type-4 (T4): Syntactically dissimilar code fragments that still share the same functionality. For example, one code fragment implementing bubble sort and another code fragment implementing quick sort are considered a Type-4 code clone pair.

As the boundary between type-3 and type-4 clones is often ambiguous, in benchmarks like BigCloneBench \cite{svajlenko2014towards} researchers further divide these two clone types into three categories: strongly type-3, moderately type-3, and weakly type-3/type-4. Each category is harder to detect than the former one. In this paper, we refer to weak type-3/type-4 clones as semantic clones.
\subsection{Graph Neural Networks}
 Traditional deep neural network models like convolutional neural network (CNN) and recurrent neural network (RNN) have shown success in Euclidean data like images or sequential data like natural language. Different from images and natural language, graph data is much more complex. An image can be seen as a set of pixels and text as a sequence of words, while in a graph, there are at least two types of information: nodes and the relationship between nodes (edges). So it is important to build novel neural network architectures for graphs.
 
The concept of GNN was first proposed in \cite{scarselli2008graph}. The target of GNN is to learn a state embedding for each node which contains the information of its neighborhood, and sometimes to learn the embedding of a whole graph. Most existing GNN models can be fit into the general framework message passing neural networks (MPNN) \cite{gilmer2017neural}, which its overall architecture is depicted in Figure 1. In the MPNN framework, a neural network model consists of two phases: message passing and readout. Suppose we have a graph $G=({\bf V}, {\bf E})$ where ${\bf V}$ is the set of vertices and ${\bf E}$ is the set of edges. Each node in $G$ retains a state ${\bf h}$, and each edge is assigned an embedding ${\bf e}$. The message passing step update the hidden state of nodes by:
\begin{align}
& {\bf m}_{j\rightarrow i}=f_{message}({\bf h}_{i}^{(t)},{\bf h}_{j}^{(t)},{\bf e}_{ij}), \forall(i,j)\in {\bf E} \\
& {\bf m}_{i}=f_{aggregate}(\{{\bf m}_{j\rightarrow i}|\forall(i,j)\in {\bf E}\}) \\
& {\bf h}_{i}^{(t+1)}=f_{update}({\bf h}_{i}^{(t)},m_{i}) 
\end{align}
Where $f_{message}$ is the message function and $f_{update}$ is the vertex update function. $f_{aggregate}$ is an aggregation function which we often use direct sum. Equations (1) and (2) can be seen as an aggregator in which each node gathers information from its neighbors. Equation (3) is an updater that updates the hidden state of all nodes \cite{zhou2018graph}. During the message passing phase, the above updating process runs for $T$ steps. In the readout phase, the model computes a vector representation for the whole graph with the readout function $f_{R}$ by:
\begin{equation}
    {\bf h}_{G}=f_{R}(\left \{ {\bf h}_{i}^{T}|i\in {\bf V} \right \})
\end{equation}

\begin{figure*}[htbp]
\centerline{\includegraphics[height=6cm, width=13cm]{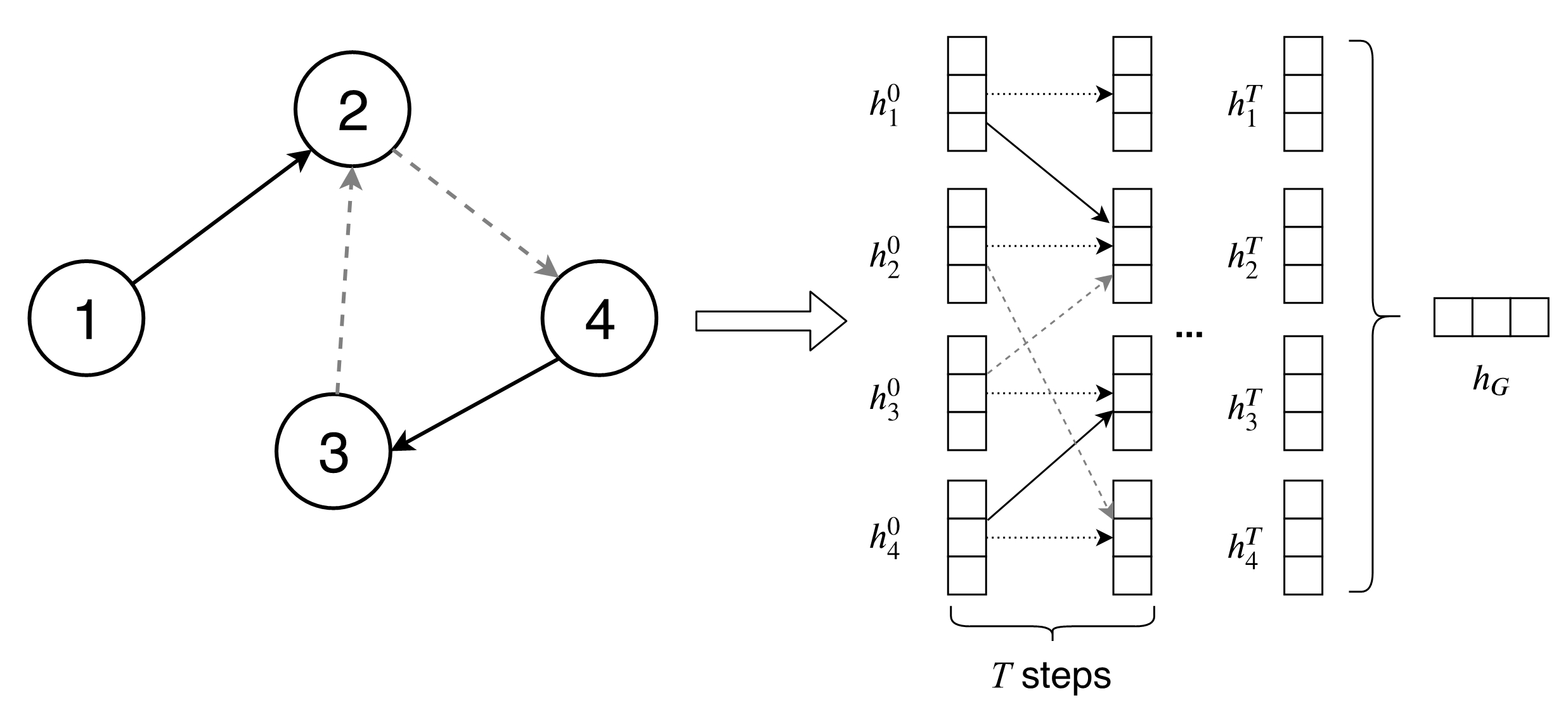}}
\caption{Example of a GNN model applied on a directed graph.}
\label{fig}
\end{figure*}

\section{Problem Definition}

Given two code fragments $C_{i}$ and $C_{j}$, we set a constant label $y_{ij}$ for them to indicate whether $(C_{i},C_{j})$ is a clone pair or not. Then for a
set of code fragments pairs with known clone labels we can build a training set $D=\left \{ (C_{i},C_{j},y_{ij})\right \}$. We aim to train a deep learning model for learning a function $\phi$ that maps a code fragment $C$ to a feature vector $v$ so that for any pair of code fragments $(C_{i},C_{j})$, their similarity score $s_{ij}=\phi (C_{i},C_{j})$ is as close to the corresponding label $y_{ij}$ .

In the inference phase, in order to determine whether a pair of code fragments $(C_{i},C_{j})$ is a clone pair, we set a threshold value $\sigma$ between true and false clone pairs. $(C_{i},C_{j})$ is a true clone pair if their similarity score $s_{ij}\geq \sigma$ and vice versa.

\begin{figure*}[htbp]
\centerline{\includegraphics[height=2cm, width=18cm]{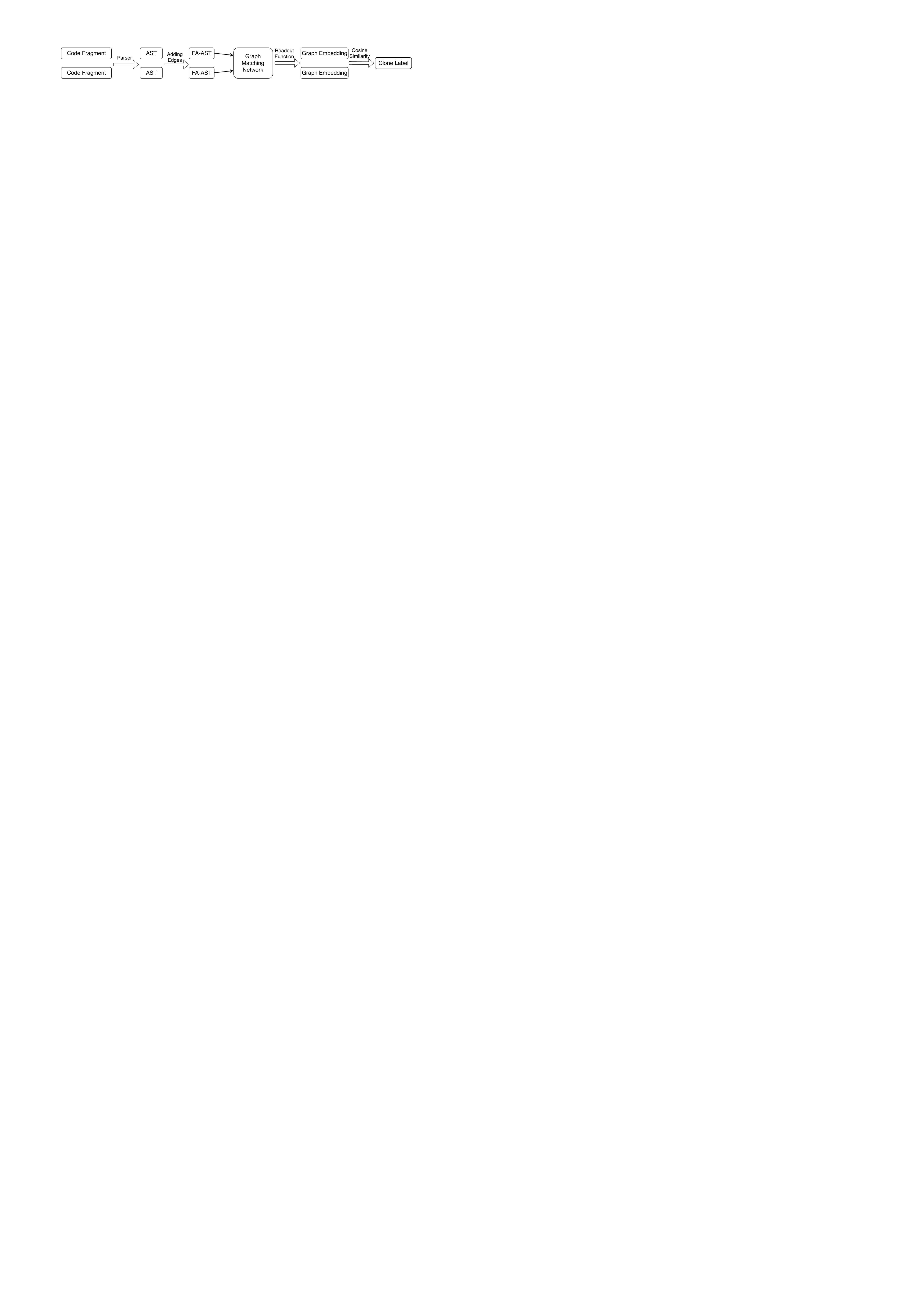}}
\caption{The overview of our approach.}
\label{fig}
\end{figure*}
\section{Proposed Approach}
In this section, we first introduce an overview of our proposed approach based on program graphs and graph neural networks. Next, we describe the process of building a graph representation: flow-augmented abstract syntax tree (FA-AST) for code fragments. We then explain the technical details of our neural network models: gated graph neural networks (GGNN) and graph matching networks (GMN).

\subsection{Approach Overview}
Figure 2 shows an overview of our approach. To process a code fragment, we first parse it into its AST. Next, we build a graph representation FA-AST for the code fragment by adding edges representing control and data flow to its AST. Then we initialize the embeddings of FA-AST nodes and edges before jointly feeding a pair of vectorized FA-ASTs into a graph matching network. The graph matching network then computes vector representation for all nodes in both FA-ASTs. To detect code clones, we use a readout function to pool the vectors of nodes into a graph-level vector representation for each FA-AST separately. After we get the vector representations of both programs, we use the cosine similarity of these two vectors to measure their similarity. If the similarity score is larger than the threshold $\sigma$, we consider the two code fragments as a clone pair. We apply the mean squared error (MSE) loss to train our model: 
\begin{equation}
    \frac{1}{d}\sum_{i=1}^{d}(y_{i}-\hat{y}_{i})^{2}
\end{equation}

Here $d$ is the dimension of $y_{i}$ and $\hat{y}_{i}$. Since in our clone detection task our prediction is a single real value as the similarity between two code snippets, so in our model $d=1$.
\subsection{Building Graphs Based on Abstract Syntax Trees}
Program ASTs only represent the syntactic structure of code, so we add different types of edges to ASTs. Although program in some languages can be converted into control flow graphs in the form of assembly language or some intermediate representations (IR), we do not directly use these control flow graphs for the following reasons:

1) For a single program, the number of edges in a control flow graph is often far fewer than in an AST. For graph neural networks, fewer edges mean less information passing between nodes, and the states of nodes are less updated.

2) Most nodes in a control flow graph is a statement expression rather than a single token. If we embed those nodes using simple approaches like bag of words, we will lose the semantic information within these nodes. Another reasonable approach is to build a sub-graph for each statement, but this will significantly increase the computational cost of our neural network model.

To extract ASTs from Java programs, we use a python package javalang $\footnote{https://github.com/c2nes/javalang}$. Below we use the node types, values, and production rules in javalang to describe Java ASTs.

To build the graph representation for programs, we construct the following types of edges based on abstract syntax trees:

\textit{Child}: connect a non-terminal AST node to each of its children according to the AST.

\textit{Parent}: connect a non-root AST node to its parent node.

\textit{NextSib}: connect a node to its next sibling (from left to right). Because graph neural networks do not consider the order of nodes, it is necessary to provide the order of children to our neural network model.

\textit{NextToken}: connect a terminal node to the next terminal node. In ASTs, terminal nodes refer to the identifier tokens in program source code, so a \textit{NextToken} edge connects an identifier token to the next token in the corresponding source code.

\textit{NextUse}: a \textit{NextUse} edge connects a node of a variable use to its next appearance. \textit{NextUse} edges can exploit useful data flow information from ASTs.

Apart from the above edge types, we add several types of edges to represent the control flow of programs. In this paper we focus on the following basic control flow types: sequential execution, If statements, While and For loops. Since other control flow structures like DoWhile statements and Case blocks appear much less in programs and are not supported by some programming languages (e.g., Python),  we omit to add control flow edges for them. We describe the details for control flow edges in FA-AST as follows:

1. {\fontfamily{\ttdefault}\selectfont If} statements: In AST, an {\fontfamily{\ttdefault}\selectfont IfStatement} node contains two or three children. The first child is the {\fontfamily{\ttdefault}\selectfont If} condition. The second (and third) children is the {\fontfamily{\ttdefault}\selectfont If} body when the condition is true (or false). As shown in figure, we add a \textit{CondTrue} edge from the condition node to the {\fontfamily{\ttdefault}\selectfont ThenStatement} node and a \textit{CondFalse} edge from the condition node to the {\fontfamily{\ttdefault}\selectfont ElseStatement} node.

\begin{figure}[htbp]
\centerline{\includegraphics[height=3cm, width=9cm]{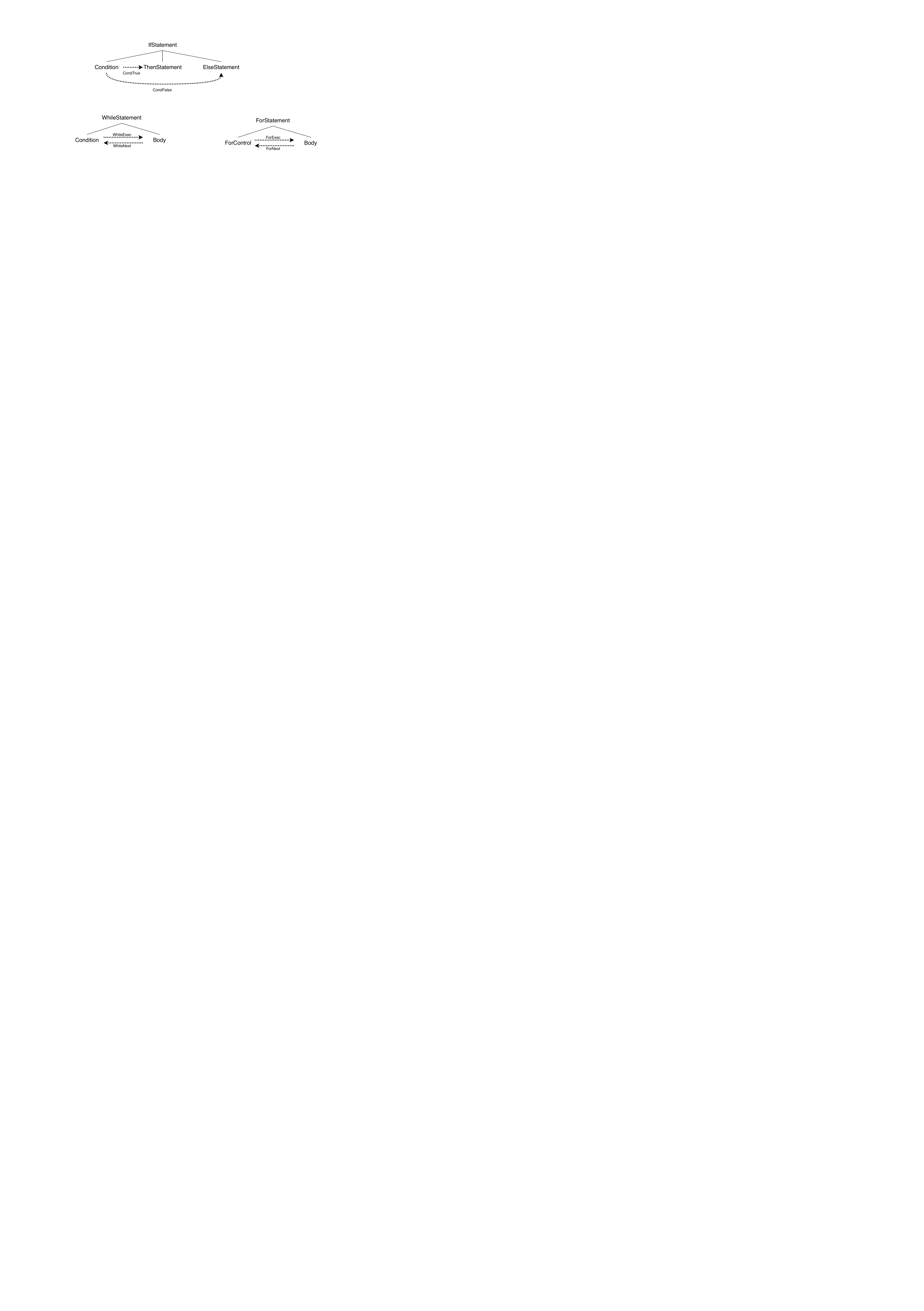}}
\caption{Control flow edges for If statements.}
\label{fig}
\end{figure}

2. {\fontfamily{\ttdefault}\selectfont While} statements: A {\fontfamily{\ttdefault}\selectfont While} node has two children: a condition code and a body node. We connect a \textit{WhileExec} edge from the condition node to the body node, and a \textit{WhileNext} node from the body node to the condition node to simulate the execution process of loops.
\begin{figure}[htbp]
\centerline{\includegraphics[height=2cm, width=6cm]{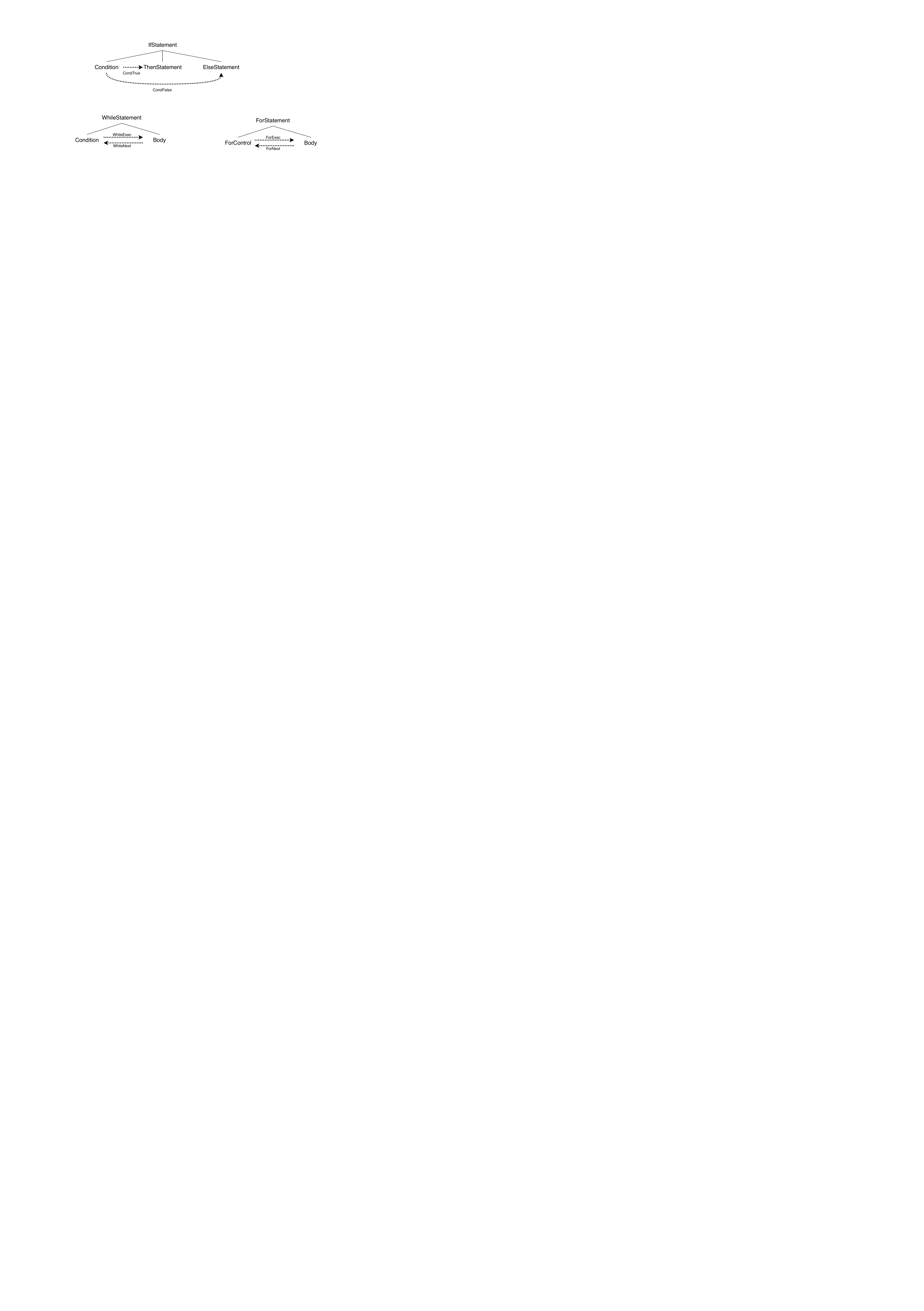}}
\caption{Control flow edges for While statements.}
\label{fig}
\end{figure}

3. {\fontfamily{\ttdefault}\selectfont For} statements: A {\fontfamily{\ttdefault}\selectfont For} node has two children: a {\fontfamily{\ttdefault}\selectfont ForControl} node and a body node. similar to the {\fontfamily{\ttdefault}\selectfont While} nodes, we add a \textit{ForExec} edge and a \textit{ForNext} edge between the two children of {\fontfamily{\ttdefault}\selectfont For} nodes. 

\begin{figure}[htbp]
\centerline{\includegraphics[height=2cm, width=6cm]{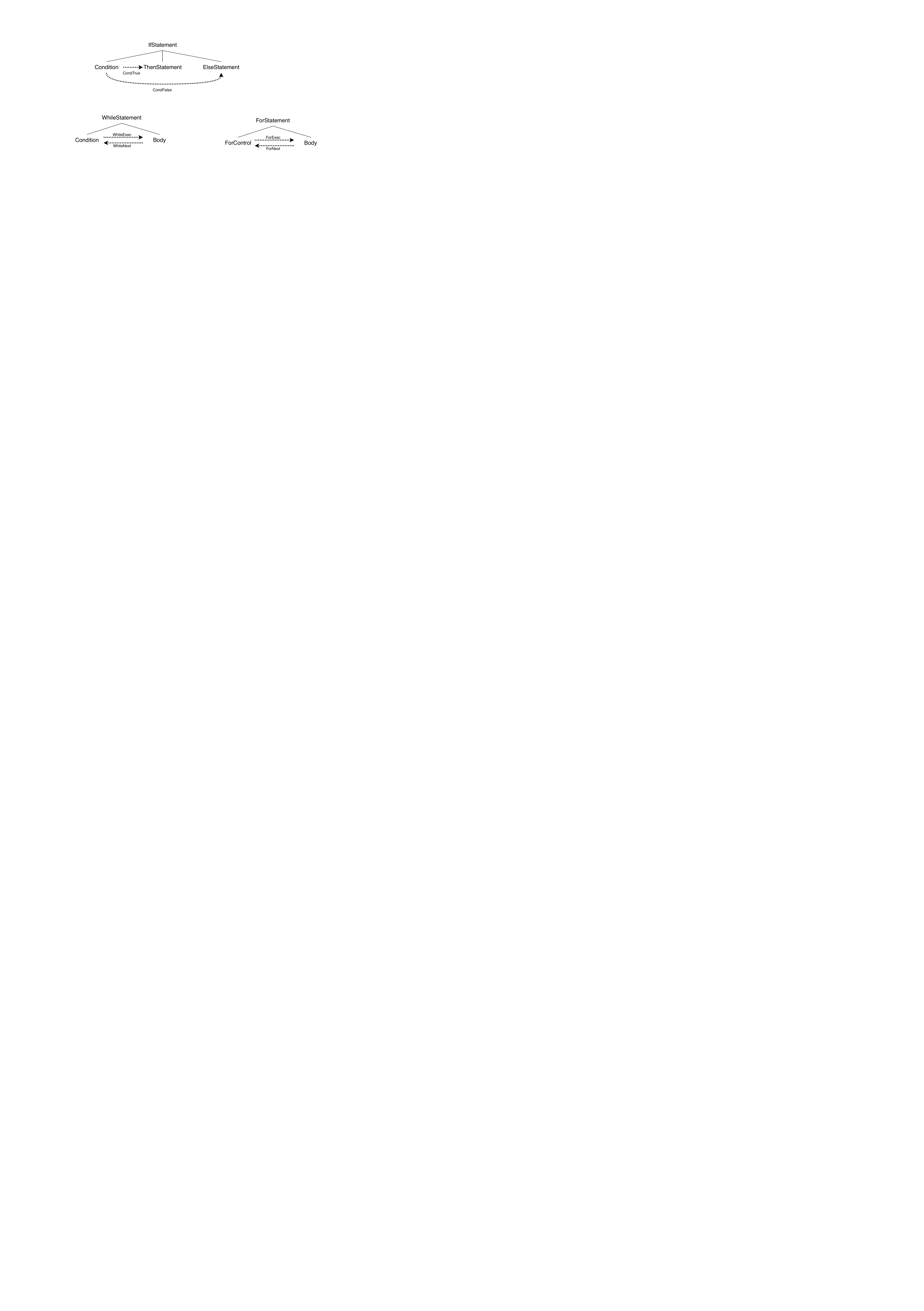}}
\caption{Control flow edges for For statements.}
\label{fig}
\end{figure}

4. Sequential execution: in Java, the sequential execution of statements exist in code clocks such as method bodies or loop bodies. A {\fontfamily{\ttdefault}\selectfont BlockStatement} node is the root of a sequence of statement AST-subtrees which are executed sequentially. Different from the control flow nodes we mentioned before, a {\fontfamily{\ttdefault}\selectfont BlockStatement} node can have an arbitrary number of children, so we add a \textit{Nextstmt} edge between the root of each statement subtree to its next sibling.
\begin{figure}[htbp]
\centerline{\includegraphics[height=2cm, width=6cm]{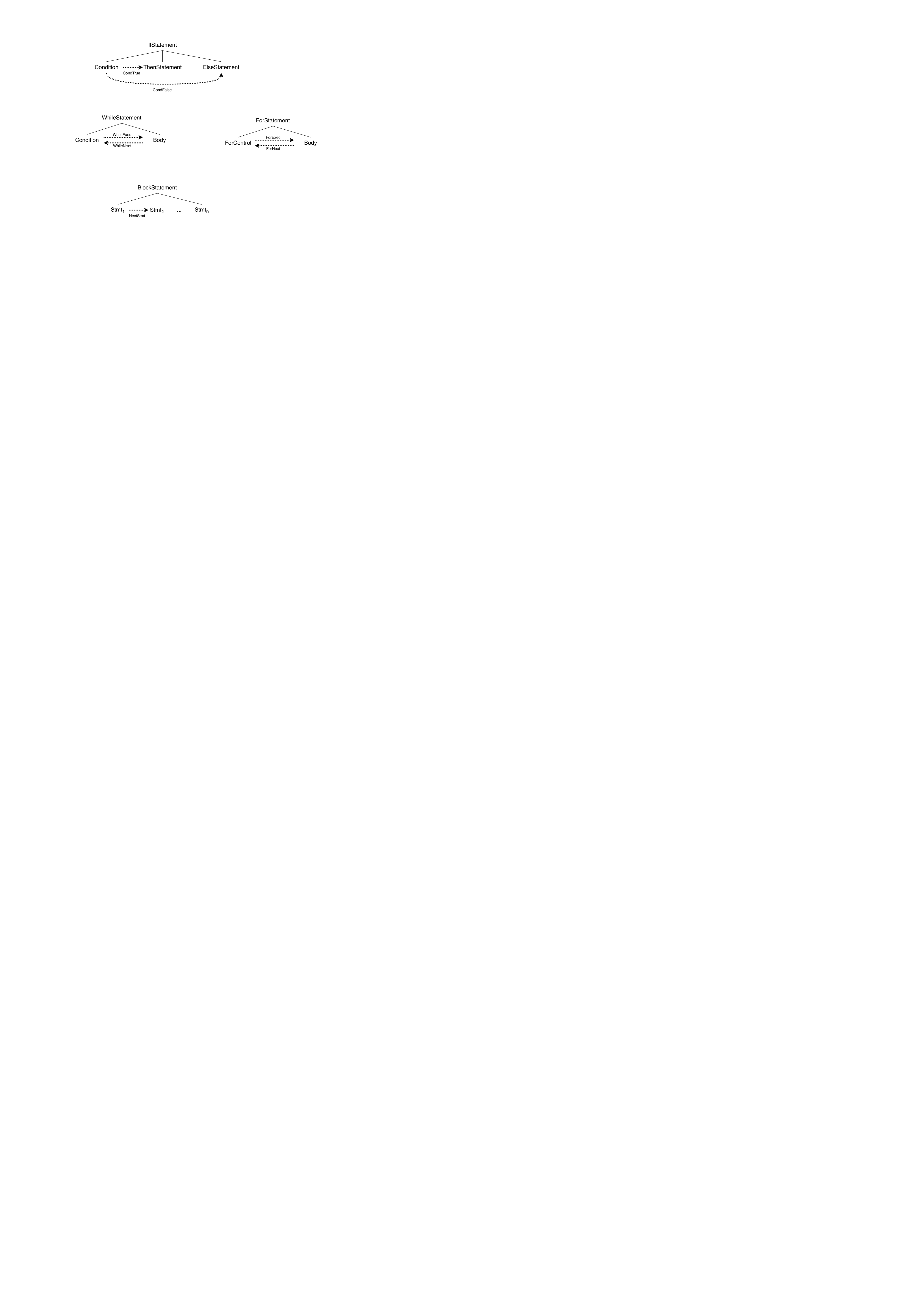}}
\caption{Control flow edges for sequences of statements.}
\label{fig}
\end{figure}

Finally, to increase the frequency of message passing, for each edge without a backward edge (e.g., \textit{CondTrue} and \textit{NextStmt}), we add an additional backward edge for them.
\subsection{Neural Network Model for Modeling Code Pairs}
In this paper, we use two different types of graph neural networks: a traditional GNN for graph embeddings and a graph matching network \cite{li2019graph} which jointly models two graphs simultaneously.
\subsubsection{Graph Embedding Model} 
In this model, we use a gated graph neural network (GGNN) \cite{li2015gated} to learn the embeddings for graphs. GGNN follows the GNN framework we introduced in section II. For GGNN, we use a multilayer perceptron (MLP) as $f_{message}$ and a gated recurrent unit (GRU) \cite{cho2014properties} as $f_{update}$. Namely, the propagation process of GGNN is:
\begin{equation}
\begin{split}
  & m_{j\rightarrow i}=\mathrm{MLP}({\bf h}_{i}^{(t)},{\bf h}_{j}^{(t)},{\bf e}_{ij}), \forall(i,j)\in {E_{1}\cup E_{2}}    \\
  & {\bf m}_{i}=\sum_{j} m_{j\rightarrow i} \\
  & {\bf h}_{i}^{(t+1)}=GRU({\bf h}_{i}^{(t)},{\bf m}_{i}) \\
  \end{split}
\end{equation}
For the readout function $f_{G}$, we follow the function proposed in \cite{li2015gated}:
\begin{equation}
    {\bf h}_{G}=\mathrm{MLP}_{G}(\sum_{i\in V}\sigma (\mathrm{MLP}_{gate}({\bf h}_{i}^{(T)}))\odot MLP({\bf h}_{i}^{(T)}))
\end{equation}

\begin{figure*}[htbp]
\centerline{\includegraphics[height=4cm, width=14cm]{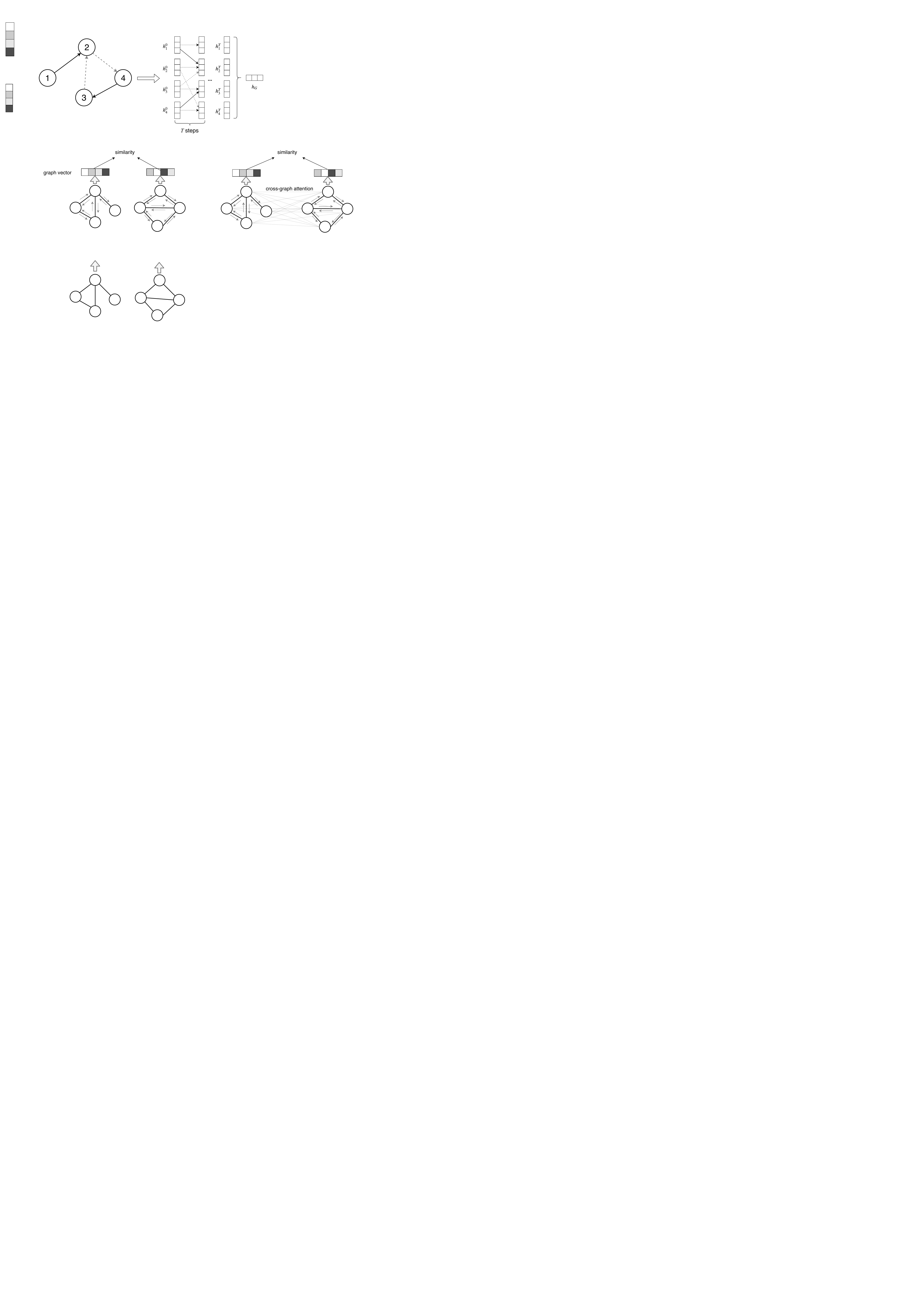}}
\caption{Basic architecture of the GGNN embedding model (left) and GMN model (right).}
\label{fig}
\end{figure*}

\subsubsection{Graph Matching Networks}
 Graph Matching Networks (GMN) framework defined by \cite{li2019graph} can jointly learn embeddings for a pair of graphs. Apart from the traditional GNN propagation process, GMN additionally computes a cross-graph attention between nodes from two graphs. Figure 7 illustrates the difference between our GGNN graph embedding approach and GMN. Although a GMN model takes two graphs as input at a time, it can still produce the separate embeddings for each input graph. The complete propagation process is demonstrated as following:

\begin{align}
& m_{j\rightarrow i}=f_{message}({\bf h}_{i}^{(t)},{\bf h}_{j}^{(t)},{\bf e}_{ij}), \forall(i,j)\in {E_{1}\cup E_{2}} \\
& \mu _{j\rightarrow i}=f_{match}({\bf h}_{i}^{(t)},{\bf h}_{j}^{(t)}), \forall i\in V_{1},j\in V_{2}\ or\ i\in V_{2},j\in V_{1} \\
& {\bf h}_{i}^{(t+1)}=f_{node}({\bf h}_{i}^{(t)},\sum_{j} m_{j\rightarrow i},\sum_{j'} \mu_{j'\rightarrow i}) \\
& {\bf h}_{G_{1}}=f_{G}(\{{\bf h}_{i}^{(T)}\}_{i\in V_{1}}) \\
& {\bf h}_{G_{2}}=f_{G}(\{{\bf h}_{i}^{(T)}\}_{i\in V_{2}})
\end{align}

Where $f_{message}$ is an MLP and $f_{match}$ is an attention mechanism defined by:
\begin{align}
& a_{j\rightarrow i}=\frac{{\rm exp}(s_{h}({\bf h}_{i}^{(t)},{\bf h}_{j}^{(t)}))}{\sum_{j'}{\rm exp}(s_{h}({\bf h}_{i}^{(t)},{\bf h}_{j'}^{(t)}))} \\
& \mu _{j\rightarrow i}=a_{j\rightarrow i}({\bf h}_{i}^{(t)}-{\bf h}_{j}^{(t)})
\end{align}

Here $s_{h}$ is a vector similarity function which we use dot product in our paper. $f_{node}$ is a GRU cell which ${\bf h}_{i}^{(t)}$ is its current hidden state at timestep $t$ and the concatenation of $\sum_{j} m_{j\rightarrow i}$ and $\sum_{j'} \mu_{j'\rightarrow i}$ is its input. Similar to GGNN, we also use the readout function in Equation (7) for GMN. With these settings, our GMN model is similar to the GGNN model, and the only difference is that GMN adds a cross-graph matching vector to the input of the updater GRU.

\section{Experiments}
\subsection{Experiment Data}
We evaluate our approach on two datasets: Google Code Jam (GCJ) \cite{gcj} and BigCloneBench \cite{svajlenko2014towards}.  The Google Code Jam \cite{gcj} is an online programming competition held annually by Google. In this paper, we use the version of the dataset collected by \cite{zhao2018deepsim}. The GCJ dataset consists of 1,669 Java files from 12 different competition problems. Each file is a Java class. As \cite{zhao2018deepsim} have inspected, very few files within a competition problem are syntactically similar, so we can assume that most code pairs from the same problem are type-4 clones. 

The second dataset BigCloneBench is a widely used large code clone benchmark that contains over 6,000,000 true clone pairs and 260,000 false clone pairs from 10 different functionalities. In BigCloneBench, each code fragment is a Java method. As the boundary between type-3 and type-4 clones is often ambiguous, type-3/type-4 clone pairs in BigCloneBench are further divided by a statement-level similarity score within [0, 1): strongly type-3 (ST3) with similarity in [0.7, 1.0), moderately type-3 (MT3) with similarity in [0.5, 0.7), and weakly type-3/type-4 (WT3/T4) with similarity in [0.0, 0.5). Table I summarizes the distribution of all clone types in BigCloneBench. Since the majority of code clone pairs are Weak Type-3/Type-4 clones, BigCloneBench is quite appropriate to be used for evaluating semantic clone detection. In our experiment, we follow the settings of in the CDLH paper \cite{wei2017supervised}, which discard code fragments without any tagged true or false clone pairs, left with 9,134 code fragments.

Table II shows the basic information about the two datasets in our experiment. Generally, since BigCloneBench contains far more code fragments than GCJ, its vocabulary size is significantly larger. On the other hand, code fragments in GCJ are usually longer than in BigCloneBench. This is mainly because each BigCloneBench code fragment only implements a single functionality like bubble sort or file copy, while in GCJ programmers are often required to solve a more complicated algorithmic problem. For both datasets, false clone pairs are much more than true clone pairs, especially in the BigCloneBench dataset. By evaluating these two datasets, we can find out the generalizability of our approach over code clones in different domains and granularities.

\begin{table}[htbp]
\caption{Percentage of different clone types in BigCloneBench}
\begin{center}
    \begin{tabular}{llllll}
    \toprule
    Clone Type & \multicolumn{1}{l}{T1} & \multicolumn{1}{l}{T2} & \multicolumn{1}{l}{ST3} & \multicolumn{1}{l}{MT3} & \multicolumn{1}{l}{WT3/T4} \\
    \midrule
    Percentage(\%) & 0.455 & 0.058 & 0.243 & 1.014 & 98.23\\
    \bottomrule
    \end{tabular}%
\label{tab1}
\end{center}
\end{table}

\begin{table}[htbp]
\caption{Basic information of two datasets}
\begin{center}
    \begin{tabular}{lll}
    \toprule
    & \multicolumn{1}{l}{GCJ} & \multicolumn{1}{l}{BigCloneBench} \\
    \midrule
    Code fragments & 1,669 & 9,134 \\
    Average lines of code & 58.79 & 32.89 \\
    Average number of nodes  & 396.98 & 241.46  \\
    Vocabulary size   & 8,033 & 77,535 \\
    True clone pairs & 275,570 & 336,498 \\
    False clone pairs & 1,116,376 & 2,080,088 \\
    \bottomrule
    \end{tabular}%
\label{tab1}
\end{center}
\end{table}

To address the importance of the control flow, we further analyze the frequency of different control flows in our datasets. Table III shows the number of occurrences for different control flow nodes in two datasets. For both datasets, {\fontfamily{\ttdefault}\selectfont BlockStatement} is the most frequent control flow, since sequential executions widely exist in nearly all programs. {\fontfamily{\ttdefault}\selectfont WhileStatement} is the fewest or second-fewest among the four control flow types we mentioned in FA-AST. An interesting difference between the two datasets is that {\fontfamily{\ttdefault}\selectfont ForStatement} appears much more times in GCJ than in BigCloneBench. This is probably because in programming contests, sometimes programmers need to implement complicated algorithms that contain a lot of {\fontfamily{\ttdefault}\selectfont For} loops. Since {\fontfamily{\ttdefault}\selectfont DoStatement} and {\fontfamily{\ttdefault}\selectfont SwitchStatement} appear much fewer than the other control flows, we decide not to add edges for these two control flows, as shown in Section IV. In general, as code fragments in GCJ are usually longer than in BigCloneBench, control flow nodes appear more in GCJ dataset.
\begin{table}[htbp]
\caption{Average occurrences of control flow nodes in our datasets.}
\begin{center}
    \begin{tabular}{lll}
    \toprule
    & \multicolumn{1}{l}{GCJ} & \multicolumn{1}{l}{BigCloneBench} \\
    \midrule
    {\fontfamily{\ttdefault}\selectfont IfStatement} & 3.114 & 2.724 \\
    {\fontfamily{\ttdefault}\selectfont WhileStatement} & 0.437 & 0.441 \\
    {\fontfamily{\ttdefault}\selectfont ForStatement}  & 4.064 & 0.422  \\
    {\fontfamily{\ttdefault}\selectfont BlockStatement}   & 7.049 & 3.274 \\
    {\fontfamily{\ttdefault}\selectfont DoStatement}   & 0.006 & 0.014 \\
    {\fontfamily{\ttdefault}\selectfont SwitchStatement}   & 0.013 & 0.012 \\
    \bottomrule
    \end{tabular}%
\label{tab1}
\end{center}
\end{table}
\subsection{Experiment Settings}

We compare our approach with the following clone detecting approaches:

DECKARD \cite{jiang2007deckard} is an AST-based clone detector which generates characteristic vectors for each AST subtree using predefined rules and then clusters them to detect code clones. 

RtvNN \cite{white2016deep} first uses an RNN language model to learn the embeddings for program tokens, then use a recursive autoencoder \cite{socher2011semi} to learn representations for ASTs. In order to represent ASTs by recursive neural networks, ASTs are turned into full binary trees. 

CDLH \cite{wei2017supervised} uses binary Tree-LSTM \cite{tai2015improved} to encode ASTs, and a hash function to optimize the distance between the vector representation of AST pairs by hamming distance.

ASTNN \cite{zhang2019novel} uses recursive neural networks to encode AST subtrees for statements, then feed the encodings of all statement trees into an RNN to compute the vector representation for a program. The similarity score between code pairs is measured by the L1 norm.

We implement the GGNN and GMN model with PyTorch $\footnote{https://pytorch.org}$ and its extension library PyTorch Geometric \cite{fey2019fast}. We set the dimension of graph neural network layers and token embeddings to 100. In both experiments, we run the GNN propagation for 4 steps. Token embeddings are initialized randomly and trained together with the model. We train our neural networks using the Adam optimizer \cite{DBLP:journals/corr/KingmaB14} with a learning rate of 0.001. We set the batch size to 32. The threshold between true and false clones are tuned by the results on the validation set. We run all experiments on a server with 32 cores of 2.1GHz CPU and an NVIDIA Titan Xp GPU. Similar to our approaches, for all neural network baselines RtvNN, CDLH, and ASTNN, we also set the hidden layer size to 100. For the rest setting of these baselines, we follow the description in their original papers or released code.

For both datasets, we split the dataset for training, validation, and test set by 8:1:1. For the BigCloneBench dataset, we use the same 9,134 code fragments from \cite{wei2017supervised}. As for both datasets, the number of false clone pairs is far more than true clone pairs, so we apply data balance on training sets. For the training sets of both tasks, we randomly downsample the false clone pairs to make the ratio between true and false pairs 1:1.

\subsection{Experiment Results}
\subsubsection{Results on Google Code Jam}
\begin{table}[htbp]
\caption{Results on the GCJ Dataset}
\begin{center}
    \begin{tabular}{lrrr}
    \toprule
    Model & \multicolumn{1}{l}{Precision} & \multicolumn{1}{l}{Recall} & \multicolumn{1}{l}{F1} \\
    \midrule
    Deckard & 0.45 & 0.44 & 0.44 \\
    RtvNN & 0.20 & 0.90 & 0.33 \\
    ASTNN  & 0.98 & 0.93 & 0.95 \\
    \midrule
    FA-AST+GGNN   & 0.96 & \textbf{1.0} & 0.97 \\
    FA-AST+GMN   & \textbf{0.99} & 0.97 & \textbf{0.98} \\
    \bottomrule
    \end{tabular}%
\label{tab1}
\end{center}
\end{table}

Table IV shows the precision, recall and F1 value of our approach on the GCJ dataset. We observe that our approach far outperforms all baselines in precision, recall and F1. By exploiting both the syntactical information in the AST and semantical information of control and data flow, our approach (FA-AST+GMN) improves the F1-score on GCJ from 0.95 (ASTNN) to 0.98.

\subsubsection{Results on BigCloneBench}
\begin{table}[htbp]
\caption{Results on the BigCloneBench Dataset}
\begin{center}
    \begin{tabular}{lrrr}
    \toprule
    Model & \multicolumn{1}{l}{Precision} & \multicolumn{1}{l}{Recall} & \multicolumn{1}{l}{F1} \\
    \midrule
    Deckard & 0.93 & 0.02 & 0.03 \\
    RtvNN & 0.95 & 0.01 & 0.01 \\
    CDLH  & 0.92 & 0.74 & 0.82 \\
    ASTNN   & 0.92 & 0.94 & 0.93 \\
    \midrule
    FA-AST+GGNN   & 0.85 & 0.90 & 0.88 \\
    FA-AST+GMN   & \textbf{0.96} & \textbf{0.94} & \textbf{0.95} \\
    \bottomrule
    \end{tabular}%
\label{tab1}
\end{center}
\end{table}

Table V shows the results of BigCloneBench. Our approach achieves much higher recall (0.94) and F1 (0.95) than most baselines. Notably, our approach outperforms ASTNN by precision and F1. 

On two tasks, the F1 of GMN models both outperform GGNN models, confirming our assumption that adding cross-graph attention in the GNN propagation process can enhance the power of the model to capture code similarities. Another noticeable phenomenon is that compared to GMN models, the recall of GGNN models is often higher than their precision scores.

To further analyze the behavioral difference between GMN and GGNN on clone detection, we make a study on the changing process of different clone metrics when we adjust the threshold similarity score between true and false clone pairs. Figure 8 shows the changing process of precision, recall, and F1 on BigCloneBench test set when we gradually change the threshold similarity score from -1 to 1. Although GGNN achieves similar recall to GMN, its precision is lower than GMN, especially when the threshold is low. This results in GGNN only achieve high F1 values in a small interval (0.5,0.75), while GMN can reach a near-best F1 in a large interval (-0.5,0.75). A small change of threshold value may significantly affect the result of GGNN models, while GMN performs more stable. After we inspect the output similarity scores of both models, we found out that for a large part of false clone pairs, their outputs of GGNN are closer to 0 rather than the ground truth label -1. This indicates that in GGNN cannot effectively distinguish dissimilar code fragments from datasets, which fits the fact that GGNN achieves recall values higher than precision on both datasets. In practice, the data distribution of the validation set and test set can be largely different, so the threshold tuned on the validation set may not suit the test set. So compared to GGNN, we believe GMN is more robust to the variation of the validation set.
\begin{figure}[htbp]
\centering
\subfigure{\includegraphics[height=4.76cm, width=6.4cm]{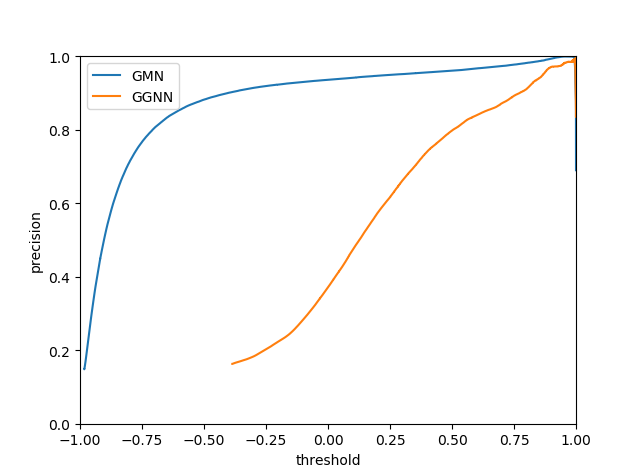}}
\subfigure{\includegraphics[height=4.76cm, width=6.4cm]{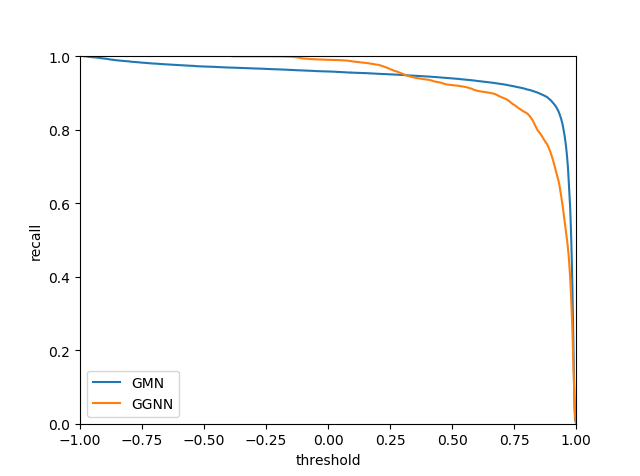}}
\subfigure{\includegraphics[height=4.76cm, width=6.4cm]{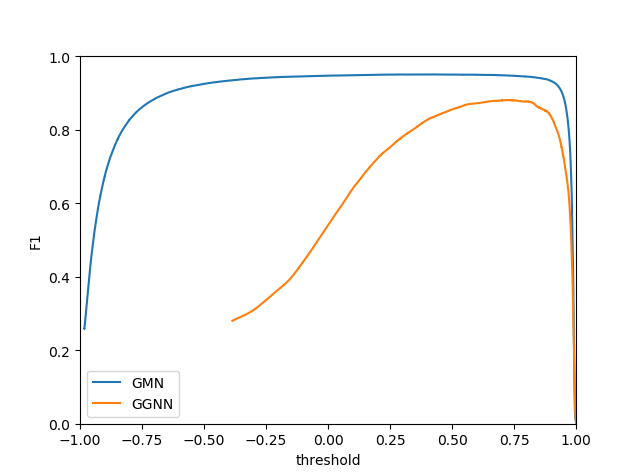}}
\caption{Precision, recall and F1 curve when changing the threshold value for BigCloneBench.}
\label{fig}
\end{figure}

\begin{figure*}[htbp]
\centerline{\includegraphics[height=8.5cm, width=13.5cm]{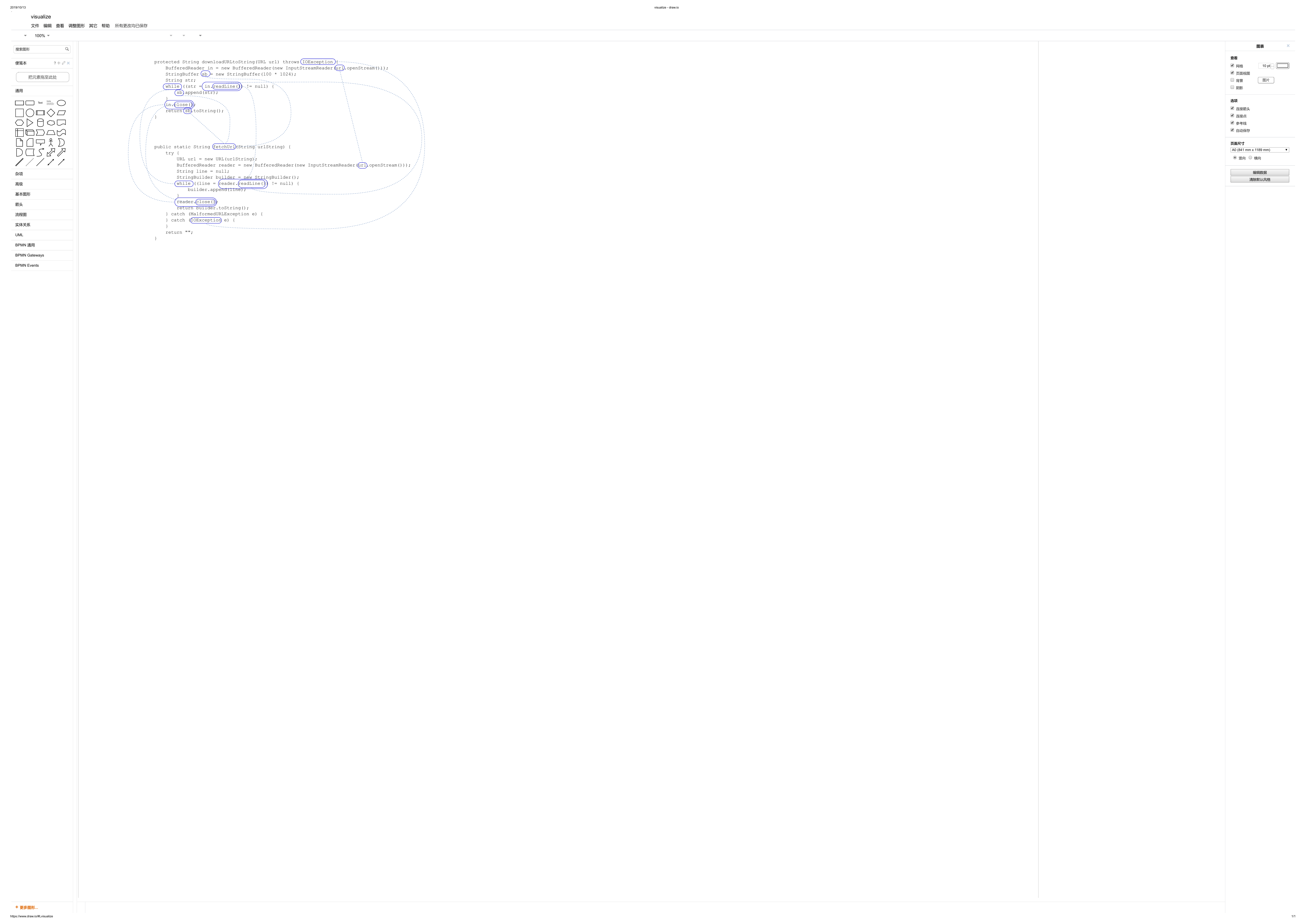}}
\caption{The corresponding locations in source code for the node pairs with the ten highest attention scores in a clone pair in BigCloneBench.}
\label{fig}
\end{figure*}

Additionally, we make a visualization study on the attention scores of the GMN model. In GMN, the cross-graph attention scores ($a_{j\rightarrow i}$ in Equation (13)) measure the similarity of two nodes from two different code fragments. After training, we assume AST node pairs with similar semantics and context should have larger attention value than other node pairs. In Figure 9, we select the ten highest attention values $a_{j\rightarrow i}$ within the whole attention matrix and display their corresponding location in their source code text. The dotted lines connect the node pairs with the highest similarities between the upper code fragment and the lower one. We can observe that GMN can learn cross graph similarities on both low-level similarities (like a method name close()) and higher-level similarities (like a While code block). Although most attention links are intuitive to human readers, there still exist a few links that cannot be well-explained (like several edges from the upper code fragment to the method name fetchUrl in the lower code fragment). This is likely because existing graph neural networks are not suited for modeling hierarchies in tree structures. As GNNs do not consider the order of neighbours, the children and parent nodes of a node are treated equally as its neighbours. Although we add directed Child and Parent edges to FA-ASTs, it still does not change the distribution of a node's neighbours.

As BigCloneBech has already labeled clone pairs with different types, we analyze the ability of our model to detect different clone types individually. Table VI shows the results on different clone types in BigCloneBench. As the results for most of our baselines are much lower than our FA-AST+GMN, here we only compare our approach (FA-AST+GMN) with ASTNN. When comparing our approach with ASTNN, we can see that our approach outperforms ASTNN in the WT3/T4 semantic clones which we concern most.
\begin{table}[htbp]
\caption{Results on different clone types in BigCloneBench}
\begin{center}
    \begin{tabular}{lrrrrrr}
    \toprule
    \multirow{2}*{Type} & \multicolumn{3}{c}{ASTNN} & \multicolumn{3}{c}{FA-AST+GMN}\\
    \cmidrule(lr){2-4} \cmidrule(lr){5-7}
     & \multicolumn{1}{l}{Precision} & \multicolumn{1}{l}{Recall} & \multicolumn{1}{l}{F1} & \multicolumn{1}{l}{Precision} & \multicolumn{1}{l}{Recall} & \multicolumn{1}{l}{F1}\\
    \midrule
    T1 & 100 & 100 & 100 & 100 & 100 & 100\\
    T2 & 100 & 100 & 100 & 100 & 100 & 100\\
    ST3  & 100 & 99.6 & 99.8 & 100 & 99.6 & 99.8\\
    MT3   & 100 & 97.9 & 98.9 & 100 & 96.5 & 98.2\\
    WT3/T4   & 93.3 & 92.2 & 92.8 & 95.7 & 93.5 & 94.6\\
    \bottomrule
    \end{tabular}%
\label{tab1}
\end{center}
\end{table}

We further draw the ROC curve of our approaches and compare them with the best baseline ASTNN. The ROC curve and ROC\_AUC score for our approaches and ASTNN on BigCloneBench are shown in Figure 10. Similar to the results shown in Table V, FA-AST achieves the highest ROC\_AUC score among the three approaches, and the result of ASTNN is a little higher than FA-AST+GGNN but lower than FA-AST+GMN.

\begin{figure}[htbp]
\centerline{\includegraphics[height=6cm, width=8cm]{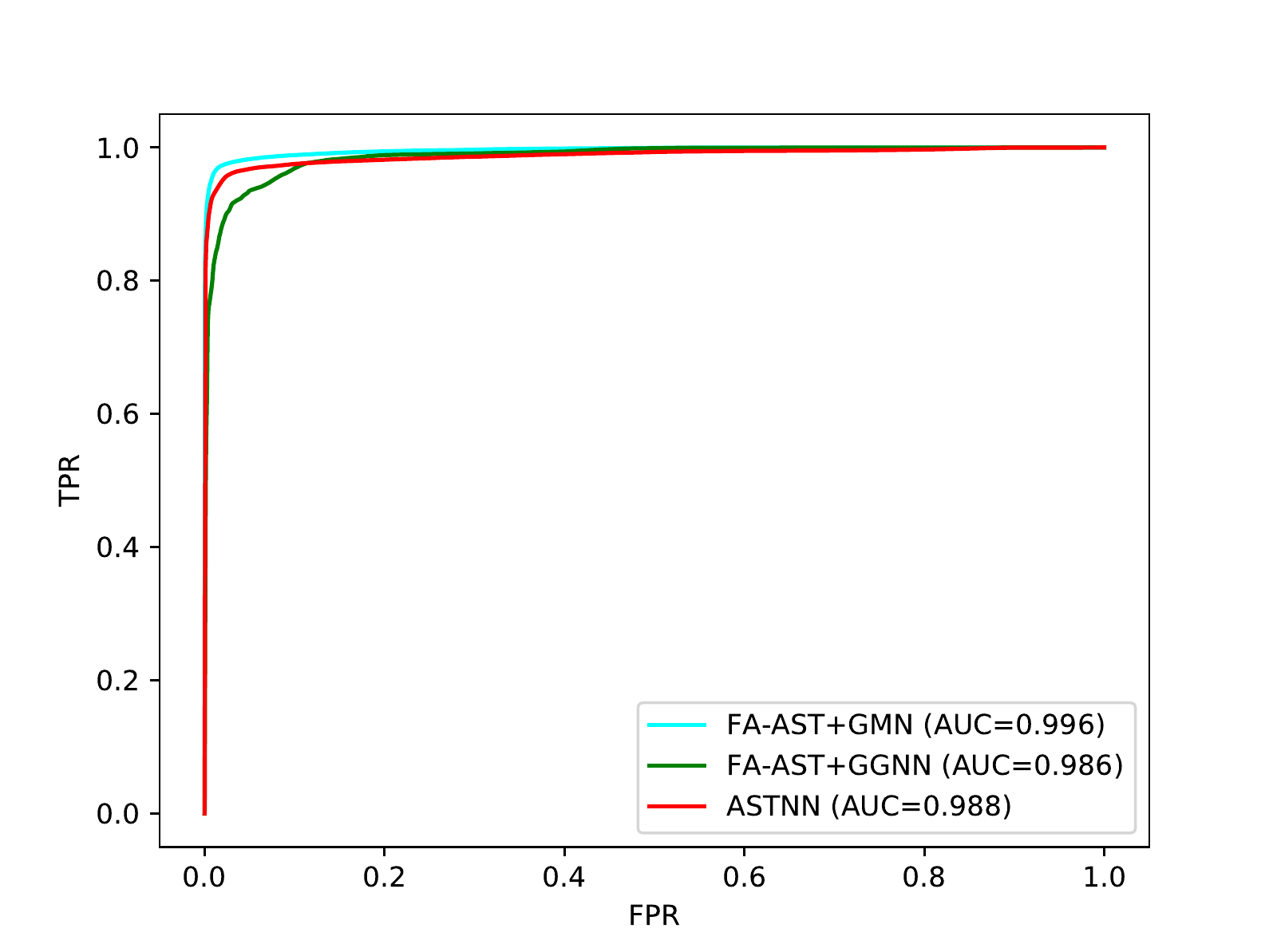}}
\caption{The ROC curve and ROC\_AUC score for FA-AST approaches and ASTNN on the test set of BigCloneBench.}
\label{fig}
\end{figure}

\section{Discussion}
In this section, we first discuss the different behaviors between our approaches and other baselines. Then we discuss some issues which our work does not solve at this point and are worth investigating in the future.

\subsection{The Advantage of Our Approach Over ASTNN}
We believe our approach outperforms previous deep learning-based clone detection approaches for the following reasons:

1) Our graph representation of programs, FA-AST, contains both syntactical information in ASTs and control and data flows in CFG, while previous approaches are based purely on either AST or CFG.

2) We treat a code fragment as a whole graph and directly input the graph in our neural network, while some previous approaches do not keep all structural information. For example, CDLH converts ASTs into binary trees before feeding it into a Tree-LSTM. DeepSim \cite{zhao2018deepsim} builds a semantic matrix for a code fragment by manually extracts several human-defined types of semantic features from CFG. ASTNN decomposes an AST into a sequence of statements subtrees by order of depth-first traverse so that it may lose the different relationships such as nesting and if/else branches between statements.

To have an intuitive view of the power of our approach over ASTNN, we demonstrate a few example true clone pairs in which our approach (FA-AST+GMN) correctly predicted while ASTNN did not. Figure 11(a) and Figure 11(b) belong to a true clone pair in BigCloneBench, in which both code fragments implement a file copy functionality. The similarity score predicted by ASTNN is 5.8e-07 (range from 0 to 1), while the similarity predicted by FA-AST is 0.94. These two code fragments are significantly different in statements, so ASTNN cannot capture the similarity between them, while GMN can learn these similarities between entire methods from training data. Figure 12 shows a false clone pair in BigCloneBench (Figure 12 (a) implements a decompress zip functionality, Figure 12 (b) implements a file copy functionality), which the similarity predicted by ASTNN is 0.94, while the similarity predicted by FA-AST is -0.27. We can see that these two code fragments are similar in both token level and statement level, so ASTNN predicted a high similarity score. From the two examples above, we believe that our approach can better capture the semantics of code fragments than ASTNN.
\begin{figure}[htbp]
\centerline{\includegraphics[height=7cm, width=8cm]{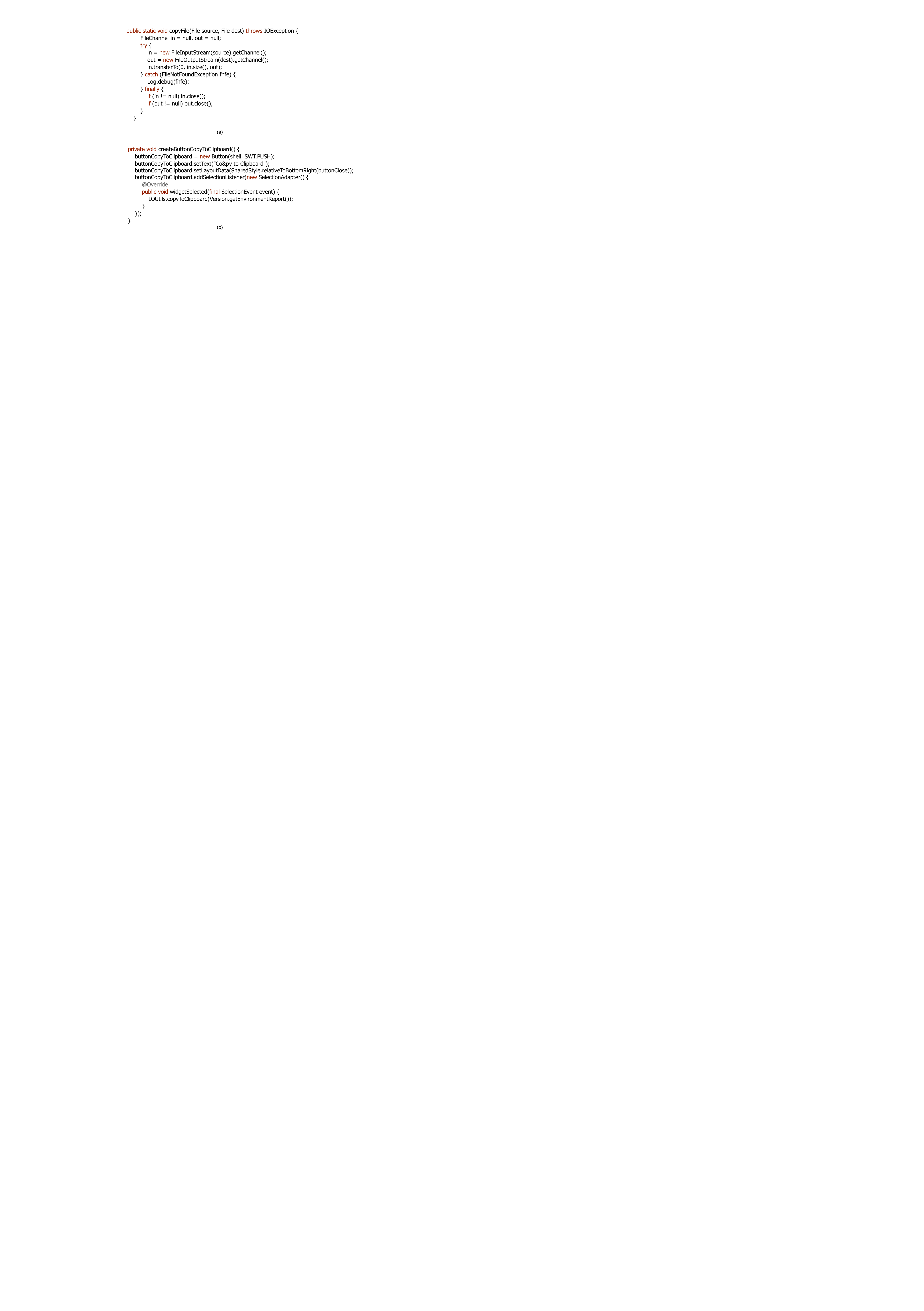}}
\caption{Example of a true clone pair in BigCloneBench which FA-AST+GMN correctly predicted as true while ASTNN wrongly predicted.}
\label{fig}
\end{figure}

\begin{figure}[htbp]
\centerline{\includegraphics[height=4.5cm, width=8cm]{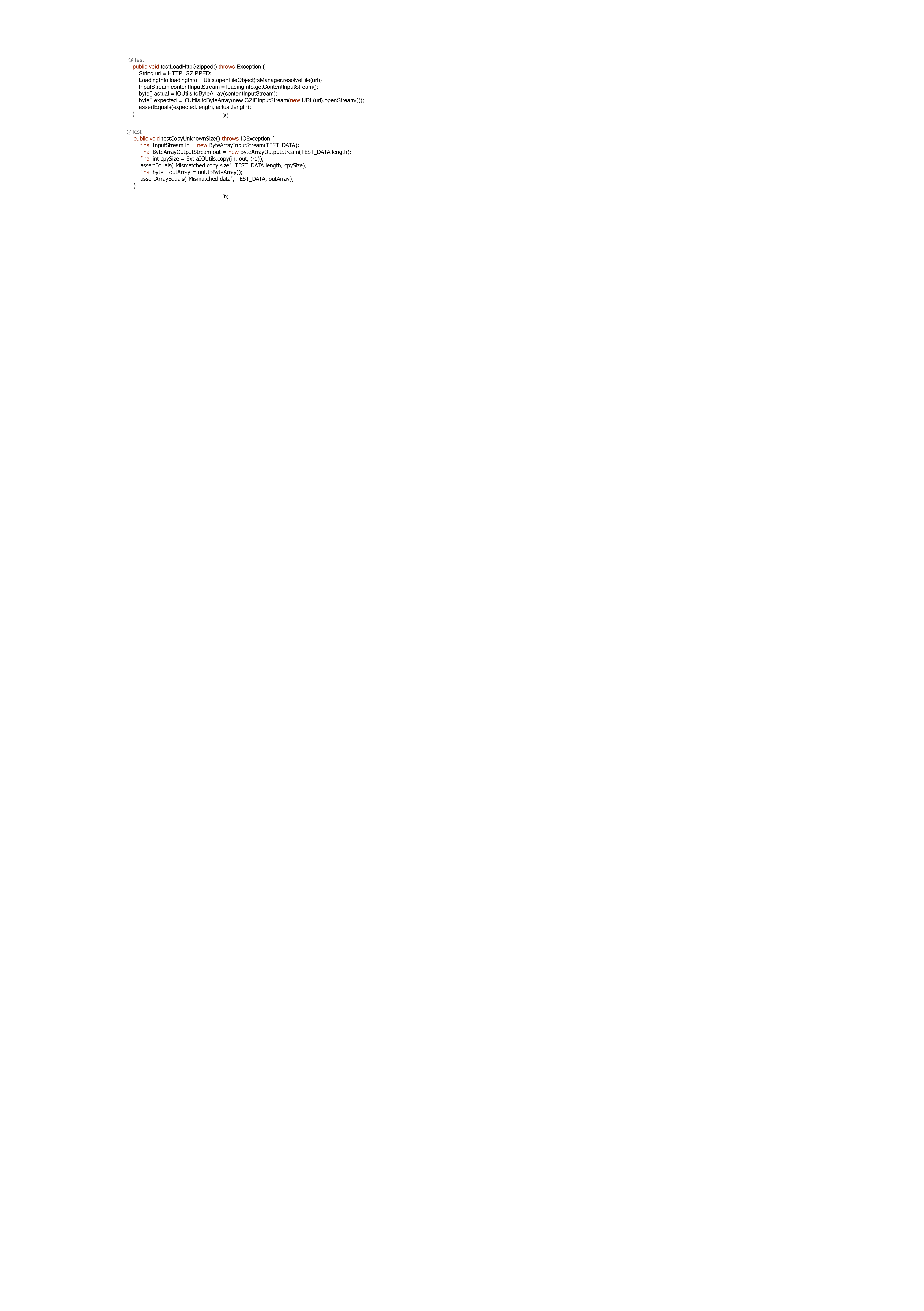}}
\caption{Example of a false clone pair in BigCloneBench which FA-AST+GMN correctly predicted as false while ASTNN wrongly predicted.}
\label{fig}
\end{figure}

\subsection{The Quality of Code Clone Datasets}
Our approaches have already shown very high results (F1 close to 1.0) on both of our tasks, but the results of ASTNN are close to ours, and the room for improvement on these two tasks are small. So we assume that some widely-used code clone datasets (like the two datasets in our paper) are not difficult enough to test the power of current deep learning models. So in the future, to test the power of up-to-date deep learning models on clone detection, we need to build larger and more complex code clone datasets. One direction is to increase the number of different functionalities in a dataset. For example, the current GCJ dataset contains 12 different functionalities, and BigCloneBench contains only ten functionalities. However, in real application cases, the code fragments are likely unable to be categorized into several classes by their functionalities. So building larger datasets with more types of different functionalities can help to test the ability of code clone detection approaches in more close-to-reality scenarios.

\subsection{Generalizability of Our Approach to Other Programming Languages}
In this paper, we use Java as an example to demonstrate the construction of FA-ASTs. In our approach, FA-AST is built with AST, control flow, and data flow, which all of them exist in most programming languages. We can follow the FA-AST building process in this paper to build graphs for other programming languages with only small modifications.

\section{Related Work}
 We introduce the related work from two perspectives: first is the application of deep learning on clone detection, and second is the application of graph neural networks on various software engineering tasks.
\subsection{Code Clone Detection with Deep Learning}
As deep learning has made a breakthrough in natural language processing, researchers have considered applying deep learning models to programming languages, which code clone detection is a well-suited task. White et al. \cite{white2016deep} used a recursive autoencoder \cite{socher2011semi} to learn representations of Java ASTs in an unsupervised manner, then used the representations to compute the similarity between code pairs. Li et al. \cite{li2017cclearner} proposed CCLearner, a purely token-based clone detector. CCLearner categorizes source code tokens into eight classes. For a pair of code fragments (methods), it calculates eight similarity scores in terms of token frequency in each category to form a feature vector that is then fed into a feedforward neural network. Wei et al. \cite{wei2017supervised} proposed CDLH, which used a hash loss to measure the similarity of two code pairs. CDLH first converted program ASTs into binary trees, then used a binary Tree-LSTM \cite{tai2015improved} to represent these trees. Wei et al. \cite{wei2018positive} proposed CDPU ( Clone Detection with Positive-Unlabeled learning), which extended CDLH with adversarial training. Different from previous supervised approaches, CDPU can be trained in a semi-supervised way using a small number of label clones and a large number of unlabeled code pairs.  Zhao et al. \cite{zhao2018deepsim} proposed a deep-learning based clone detection framework DeepSim. Different from other deep-learning based clone detection techniques, the inputs of DeepSim is not code fragments, but semantic matrices with manually extracted semantic features from CFGs. Although DeepSim first applies deep learning approach on the Google Code Jam dataset and achieved the previous state-of-the-art, we do not compare it with our approach because we cannot reproduce their experiments from the code they released. Saini et al. \cite{saini2018oreo} proposed Oreo, which uses a Siamese Network consists of two feedforward networks to predict code clones. The inputs of the neural network are a series of human-defined software metrics. They trained Oreo using 50k Java projects from GitHub and evaluate their approach on BigCloneBench. Zhang et al. \cite{zhang2019novel} proposed a program representation model ASTNN, which aimed to mitigate the long dependency problem in previous sequential models. The authors evaluated their model on code classification and clone detection.
\subsection{Graph Neural Networks for Software Engineering}
Li et al. \cite{li2015gated} proposed gated graph neural network (GGNN) which used a GRU cell to update the state of nodes. They evaluated their model on a simple program verification task to detect null pointers. The input they used is not the entire program but the memory heap states of programs. Other works try to apply GNNs on entire code fragments. To represent a program with a graph, one straightforward approach is to use control flow graphs\cite{phan2017convolutional,li2019graph}. Phan et al. \cite{phan2017convolutional} used graph convolutional network for defect detection on control flow graphs in C. To produce CFGs for C, they first compiled C source code to assembly code, then generated CFGs from the compiled assembly code. Li et al. \cite{li2019graph} proposed graph matching networks (GMN) for learning the similarity between two graphs. They applied their model to compute the similarity between control flow graphs of binary functions. Another group of works tries to create program graphs using AST \cite{allamanis2017learning,brockschmidt2018generative}. Allamanis et al. \cite{allamanis2017learning} used GGNN to learn representations for C\# programs for two tasks: variable naming and correcting variable misuse. Brockschmidt et al. \cite{brockschmidt2018generative} used GGNN to generate program expressions for code completion in C\#.

\section{Conclusion and Future Work}
Code clone detection has been a widely-studied field in software engineering, but few existing approaches can effectively detect semantic clones (i.e., clones that are very different syntactically). In this paper, we propose a novel approach that leverages explicit control and data flow information for code clone detection. Our approach applies two different GNNs, gate graph neural networks and graph matching networks over a flow-augmented AST (FA-AST). By building FA-AST using original ASTs and flow edges, our approach can directly capture the syntax and semantic structure in ASTs. Experimental results on two datasets (Google Code Jam and BigCloneBench) show that by combining graph neural networks and control/data flow information, we can enhance the performance of detecting semantic code clones.

In the future, we plan to improve our neural model and explore other program representation forms to capture more accurate syntactic and semantic features of source code. Another feasible extension to our existing work is to combine ASTs with other program structures, like token sequences or data dependence graphs.

\section*{Acknowledgment}
This research is supported by the National Key R\&D Program under Grant No. 2018YFB1003904, and the National Natural Science Foundation of China under Grant No. 61832009.
\bibliographystyle{IEEEtran}
\bibliography{ref}

\end{document}